\documentclass{aa}  

\usepackage{natbib}
\bibpunct{(}{)}{;}{a}{}{,} % to follow the A&A style

\usepackage{amsmath, siunitx}
\usepackage{comment}
\usepackage{float}
\usepackage{graphicx}
\usepackage{txfonts}
\usepackage{color}
\usepackage{subcaption} % in preamble
\usepackage[colorlinks]{hyperref}

\hypersetup{
    colorlinks = true,
    linkcolor = {blue},
    citecolor = {blue},
    urlcolor = {black}
}

\usepackage{array}
\usepackage{multirow}
\newcolumntype{P}[1]{>{\centering\arraybackslash}p{#1}}
\newcolumntype{D}{ >{\centering\arraybackslash} c{1cm} }
\usepackage[table, dvipsnames]{xcolor}
\usepackage{multirow}
\definecolor{AS}{RGB}{201, 123, 132}
\definecolor{tablecol}{RGB}{252, 244, 215}
\usepackage[table]{xcolor}
\usepackage{booktabs}

\DeclareSIUnit\au{\astronomicalunit}
\DeclareSIUnit\parsec{pc}
\DeclareSIUnit\msun{\ensuremath{M_\odot}}
\DeclareSIUnit\rsun{\ensuremath{R_\odot}}
\DeclareSIUnit\lsun{\ensuremath{L_\odot}}
\DeclareSIUnit\kmpers{km/s}
\DeclareSIUnit\mJy{mJy}
\DeclareSIUnit\K{K}
\DeclareSIUnit\overcmsq{cm^\ensuremath{-2}}

\newcommand{\micron}{ \textmu m}

\usepackage{chemmacros}

\begin{document} 

    \title{MINDS: Complementary inclinations in the binary system HK Tau reveal gas- and ice-phase chemistry}

   \author{
	Alice Somigliana\inst{\ref{inst_MPIA}} \and
	Giulia Perotti\inst{\ref{inst_CPH},\ref{inst_MPIA}} \and
    Nicolás T. Kurtovic\inst{\ref{inst_MPIA},\ref{inst_MPE}} \and
    Thomas Henning\inst{\ref{inst_MPIA}} \and
    Myriam Benisty\inst{\ref{inst_MPIA}} \and \\
    Andrew D. Sellek\inst{\ref{inst_LEID}} \and
    Melissa McClure\inst{\ref{inst_LEID}} \and
    Zak L. Smith\inst{\ref{inst_LEID}} \and
    Aditya M. Arabhavi\inst{\ref{inst_GRN}}  \and
    Alessio Caratti o Garatti\inst{\ref{inst_INAF}} \and \\
    Valentin Christiaens\inst{\ref{inst_LEU},\ref{inst_LIE}}  \and
    Ewine F. van Dishoeck\inst{\ref{inst_LEID},\ref{inst_MPE}} \and 
    Danny Gasman\inst{\ref{inst_MPIA}}  \and
    Sierra L. Grant\inst{\ref{inst_CRNG}}  \and
    Manuel G\"udel\inst{\ref{inst_VIE},\ref{inst_ZRC}} \and \\
    Till Kaeufer\inst{\ref{inst_EXE}}  \and
    Inga Kamp\inst{\ref{inst_GRN}} \and 
    Lucas Stapper\inst{\ref{inst_MPIA}} \and
    Benoît Tabone\inst{\ref{inst_SACL}} \and
    Milou Temmink\inst{\ref{inst_LEID}} \and
    Marissa Vlasblom\inst{\ref{inst_LEID}}
    }
    
    \institute{
	Max-Planck-Institut f\"{u}r Astronomie, K\"{o}nigstuhl 17, 69117 Heidelberg, Germany\label{inst_MPIA}\\ \email{alsomigliana@mpia.de} \and      
	Niels Bohr Institute, University of Copenhagen, NBB BA2, Jagtvej 155A, 2200 Copenhagen, Denmark\label{inst_CPH} \and
    Max-Planck-Institut f\"{u}r Extraterrestrische Physik, Giessenbachstrasse 1, D-85748 Garching, Germany\label{inst_MPE} \and
    Leiden Observatory, Leiden University, PO Box 9513, 2300 RA Leiden, the Netherlands\label{inst_LEID} \and
    Kapteyn Astronomical Institute, Rijksuniversiteit Groningen, Postbus 800, 9700AV Groningen, The Netherlands\label{inst_GRN} \and
    INAF – Osservatorio Astronomico di Capodimonte, Salita Moiariello 16, 80131 Napoli, Italy\label{inst_INAF} \and
    Institute of Astronomy, KU Leuven, Celestijnenlaan 200D, 3001 Leuven, Belgium\label{inst_LEU} \and
    STAR Institute, Universit\'e de Li\`ege, All\'ee du Six Ao\^ut 19c, 4000 Li\`ege, Belgium\label{inst_LIE} \and
    Earth and Planets Laboratory, Carnegie Institution for Science, 5241 Broad Branch Road, NW, Washington, DC 20015, USA\label{inst_CRNG} \and
    Dept. of Astrophysics, University of Vienna, T\"urkenschanzstr. 17, A-1180 Vienna, Austria\label{inst_VIE} \and
    ETH Z\"urich, Institute for Particle Physics and Astrophysics, Wolfgang-Pauli-Str. 27, 8093 Z\"urich, Switzerland\label{inst_ZRC} \and
    Department of Physics and Astronomy, University of Exeter, Exeter EX4 4QL, UK\label{inst_EXE} \and
    Université Paris-Saclay, CNRS, Institut d'Astrophysique Spatiale, 91405 Orsay, France\label{inst_SACL}
    }

   \date{Received 2 March 2026 / Accepted 16 June 2026}
 
  \abstract{

    HK Tau is a roughly equal mass pre-main sequence binary system consisting of a low-inclination primary (57$^\circ$) and an edge-on (\ang{83}) secondary. We present JWST Mid-Infrared Instrument (MIRI) observations targeting both sources, taken as part of the JWST GTO program MINDS. The mid-infrared spectra reveal a line-rich, \ch{CO2}-dominated primary and a line-poor secondary; this evidence, albeit in line with the evolutionary-motivated trend uncovered by recent observations of binaries at MIRI wavelengths, is likely due to the different configuration of the two sources. Indeed, thermochemical disc models coupled with radiative transfer show that, at inclinations comparable to that of HK Tau B, only ionised atomic lines are expected to remain visible in the spectra. While blocking molecular emission lines, however, the edge-on configuration allows ice absorption bands to be visible against the continuum; in this framework, the HK Tau system provides an unprecedented opportunity to have a simultaneous view of the solid and gaseous component of a pair of coeval protoplanetary discs, thanks to the complementary inclination of the two sources.
    We detect water ice at 6.2 and 13.6\micron, \ch{CO2} ice at 15.2\micron, and \ch{NH4+} ice at 6.85\micron\ in the spectrum of HK Tau B; an additional absorption band between 8.3 and 9\micron\ is compatible with both silicate stretching and C-H bending. Neither the primary nor the secondary show signs of Polycyclic Aromatic Hydrocarbons (PAHs). Extended \ch{H2} emission is present around both sources, although much more elongated in HK Tau B. The distinctive 'X' shape centred at the location of B, combined with the intensity, morphology, and spectral characteristics of the ionised atomic lines [Ar II], [Ne II], and [Ne III] suggests a low-velocity wind origin with a wide ($\sim$ \ang{70}) semi-opening angle. The lower forbidden line fluxes and smaller spatial extent of the \ch{H2} emission around A imply that, if a wind is launched from the primary as well, it is too cold or dense to be ionised and brightly emitting.   
    }


   \titlerunning{MINDS. The HK Tau system}
   \authorrunning{Somigliana, A. et al.}
    
   \maketitle

\section{Introduction}

The majority of stars form in binary or higher multiplicity stellar systems \citep{offner2023origin-d5b}. The dynamical interaction between stars and discs in multiple systems has a major impact on the presence and evolution of the discs themselves, which in turn determines the budget and potential for planet formation. Theoretical studies predict mechanisms such as outer disc truncation, inner disc warping, and material ejection \citep{papaloizou1977tidal-c4c, artymowicz1994dynamics-23e, kuruwita2023contribution-4bf} to be common in multiple systems; in the last decade, facilities such as the Atacama Large Millimeter Array (ALMA) and Very Large Telescope (VLT) have provided observational confirmation of these dynamical interactions \citep{akeson2019resolved-3ac, manara2019observational-4ed, tobin2020vlaalma-1b4, rota2022observational-f65, Zagaria2022binaries}.

At infrared wavelengths, \textit{Spitzer} Space Telescope \citep{werner2004spitzer-b59} surveys have targeted $\sim 90 \%$ of the star-forming regions within 500 pc of the Sun \citep{evans2009spitzer-118}, obtaining mid-infrared spectra with the InfraRed Spectrograph (IRS) for over 2000 young stellar objects  \citep{kessler-silacci2007probing-d6e, evans2009spitzer-118, furlan2009disk-4a9, oliveira2010spitzer-12c}. These early studies put constraints on the disc frequency and lifetimes \citep{bouwman2006binarity-ba8}, dust growth and settling \citep{furlan2006survey-4d1, furlan2009disk-4a9}, and spectral properties. The latter have allowed to constrain molecular inventories through the detection of several lines (\ch{C2H2}, \ch{HCN}, \ch{CO2}, \ch{OH} and \ch{H2O} - \citealt{lahuis2006hot-f64, carr2008organic-95f, salyk2008h2o-305, pontoppidan2010spitzer-ec6}); furthermore, the detection of [Ne II] emission \citep{pascucci2007detection-166, lahuis2007c2d-e8b, najita2010spitzer-bd0, gdel2010origin-13f, espaillat2013tracing-742} has paved the way to direct infrared observations of jets \citep{gdel2010origin-13f} and photoevaporative winds (\citealt{pascucci2009evidence-e13}; see \citealt{pascucci2023role-55b} for a review). However, the spatial resolution of \textit{Spitzer}/IRS did not allow separating the contribution of the individual stars and discs in multiple stellar systems - which ultimately resulted in the spectrum of multiples being dominated by the primary. Since the advent of the Mid-InfraRed Instrument (MIRI, \citealt{rieke2015mid-infrared-6f6, wright2015mid-infrared-88f, wright2023mid-infrared-917}) on the James Webb Space Telescope (JWST, \citealt{rigby2023science-6f7}), high enough spatial resolution and sensitivity is available to separate the spectra of multiple stellar systems in the mid-infrared wavelength regime. The first JWST/MIRI analysis of a multiple system in the Medium Resolution Spectrometer mode (MRS, 4.9-27.9\micron, \citealt{wells2015mid-infrared-b57, argyriou2023jwst-f58}) was performed on DF Tau \citep{grant2024minds-a9e}, although the small separation on the sky plane (around 70 mas, 10 au) only allowed recovery of a combined spectrum for the two sources. The sample of binary systems observed with JWST/MIRI-MRS was then expanded by \cite{arulanantham2025jdisc-9a4}, with the addition of AS 205 N and S, and \cite{kurtovic2026minds-5c9}, who targeted  VW Cha, WX Cha, and RW Aur; as the separation of these systems is larger than the angular resolution of the instrument at short wavelengths, the emission from the primary and secondary component could be disentangled. Interestingly, the spectra showed dramatic differences with the primary stars being water-rich and the secondaries line-poor, at least at the current sensitivity. These findings may be a consequence of the increased accretion and radial drift resulting from dynamical disc truncation, and calls for a larger sample of binary systems to assess the robustness of the trend.

Along with molecular gas composition, ice spectroscopy is a key focus of infrared observations. Ice studies with \textit{Spitzer} mostly targeted embedded Class 0/I protostars, probing ices in protostellar envelopes \citep{pontoppidan2008c2d-4b1, boogert2008c2d-f35, berg2011spitzer-41e}. In the case of Class II systems, where the surrounding envelope has completely dissipated, ice absorption features are primarily observable in the edge-on configuration: at high inclinations (> \ang{75}), the disc itself blocks the light coming from the central star and bright inner region, which improves the contrast between the ice species and the disc continuum. Observing edge-on discs in the mid-infrared, however, requires high sensitivity, which was limited to a few sources in the \textit{Spitzer} era (such as in \citealt{pontoppidan2005ices-411, pontoppidan2007deep-362}). JWST enabled the characterisation of ices in edge-on Class II systems: MIRI/MRS observed Tau 042021 \citep{arulanantham2024jwst-ecf}, HH 48 NE \citep{sturm2024jwstmiri-3fe}, and d216-0939 \citep{potapov2025simple-6eb}, which show \ch{H2O} and \ch{CO2} ice features - the latter also \ch{NH3} and tentatively \ch{CH4} and \ch{NH4+}; furthermore, \ch{H2O} and \ch{CO} ice bands in the JWST/NIRSpec \citep{sturm2023jwst-545, pascucci2025nested-cf6} and JWST/NIRCam \citep{ballering2025water-e2e} have been detected in edge-on Class II discs. Recently, \cite{bergner2026jwst-00e} compiled NIRSpec and MIRI observations of five additional targets, detecting \ch{H2O}, \ch{CO2}, and \ch{CO} ice in all of them.

\begin{table}[ht]
\caption{Properties of the HK Tau system.}
\centering

\label{tab:properties_of_HKTau}

\begin{tabular}{l c c}
\multicolumn{3}{c}{Entire system} \\
\toprule
Parameter & Value & References \\
\midrule
Separation & 2.34\si{\arcsecond}, $\sim 300$ au & (1) \\
Distance & 128 $\pm$ 4 pc & (2), (3) \\
\bottomrule
\end{tabular}

\vspace{0.4cm}

\begin{tabular}{l c c c}
\multicolumn{4}{c}{Single components} \\
\toprule
Parameter & A & B & References \\
\midrule
Inclination & \ang{56.9}$^{+0.5}_{-0.5}$ & \ang{83.2}$^{+0.2}_{-0.2}$ & (4) \\
Position angle & \ang{175;} & \ang{41;} & (5), (6) \\
Stellar mass & $0.44^{+0.14}_{-0.11}$ & $0.37^{+0.2}_{-0.2}$ M$_{\odot}$ & (4) \\
Spectral type & M1 & M2 & (7) \\
Stellar luminosity & $0.55^{+0.15}_{-0.15}$ & $\geq 0.03^{+0.01}_{-0.01}$ L$_{\odot}$ & (8) \\
Stellar radius $\times \sin(i)$ & $1.40^{+0.1}_{-0.1}$ \si{\rsun} & -- & (9) \\
Systemic velocity & $5.98^{+0.09}_{-0.51}$ & $6.05^{+0.7}_{-0.37}$ \si{\kmpers} & (10) \\
\bottomrule
\end{tabular}

\tablefoot{\textbf{References:} (1) \cite{white2001observational-9a3} (2) \cite{gaiacollaboration2021gaia-99b}, (3) \cite{luhman2023census-289}; note that the distance is estimated from the primary, (4) \cite{manara2019observational-4ed}, (5) \cite{leinert1993systematic-b60}, (6) \cite{villenave2020observations-5b8}, (7) \cite{monin1998using-5c7}, (8) \cite{simon2019masses-8f4}; note that the luminosity of B is likely underestimated because of its inclination, (9) \cite{nofi2021projected-a87}, (10) \cite{braun2021dynamical-e59}; note that the systemic velocity is measured from ALMA data in Kinematic Local Standard of Rest (LSRK) reference.}

\end{table}

In this paper, we present JWST/MIRI-MRS observations of the pre-main sequence binary system HK Tauri (HK Tau), a $\sim 0.4$ $\rm{M}_{\odot}$ roughly equal mass binary, taken as part of the MIRI Mid-Infrared Disk Survey (MINDS, \citealt{henning2024minds-d7e}). HK Tau is located in the L1529/B18 molecular cloud, in the Taurus star-forming region (distance 128 $\pm$ \SI{4}{\parsec}, \citealt{gaiacollaboration2021gaia-99b, luhman2023census-289}). It is a well-characterised system, observed with several facilities (Hubble Space Telescope by \citealt{koresko1998circumstellar-281, stapelfeldt1998edge-on-216}; Plateau de Bure by \citealt{duchne2003layered-cf7}; Keck by \citealt{mccabe2003first-f44}; Subaru by \citealt{terada2007detection-814}; VLT by \citealt{appenzeller2005edge-on-c93, mccabe2011spatially-5ef}; \textit{Spitzer} by \citealt{furlan2006survey-4d1}; AKARI by \citealt{aikawa2012akari-f20}; ALMA by \citealt{villenave2020observations-5b8}): we summarise its main properties in Table \ref{tab:properties_of_HKTau}.

Interestingly, the disc around HK Tau A has a smaller radius than that around HK Tau B, despite the disc mass being larger \citep{duchne2003layered-cf7}; this implies that the disc sizes were established during the formation of the binary system and not the subsequent dynamical evolution \citep{artymowicz1994dynamics-23e, kuffmeier2020ionization-bef}. While the primary HK Tau A is a classical T Tauri star surrounded by a disc seen at an inclination of $\sim$ \ang{60;} \citep{manara2019observational-4ed}, HK Tau B offers a nearly edge-on view of its disc ($> \ang{83;}$; \citealt{stapelfeldt1998edge-on-216, mccabe2011spatially-5ef, villenave2020observations-5b8}) and is therefore much fainter at visible and near-infrared wavelengths. Although the two stars have very similar spectral types (we adopt the M1 and M2 classification, for primary and secondary respectively, of \citealt{monin1998using-5c7}), the edge-on disc of HK Tau B blocks the light coming from the central star \citep{stapelfeldt1998edge-on-216} and only allows scattered photons to reach the observer. The high inclination gives a direct insight on the disc vertical structure, which has been shown to be stratified, with the larger dust closer to the midplane \citep{duchne2003layered-cf7}. The binary nature of the system, with similar spectral types but complementary inclinations of the two sources, offers a unique opportunity to explore the disc chemistry in the gaseous and icy reservoirs of a pair of coeval sources with a common formation history.

\begin{figure*}[ht]
    \centering
    \includegraphics[width=0.99\linewidth]{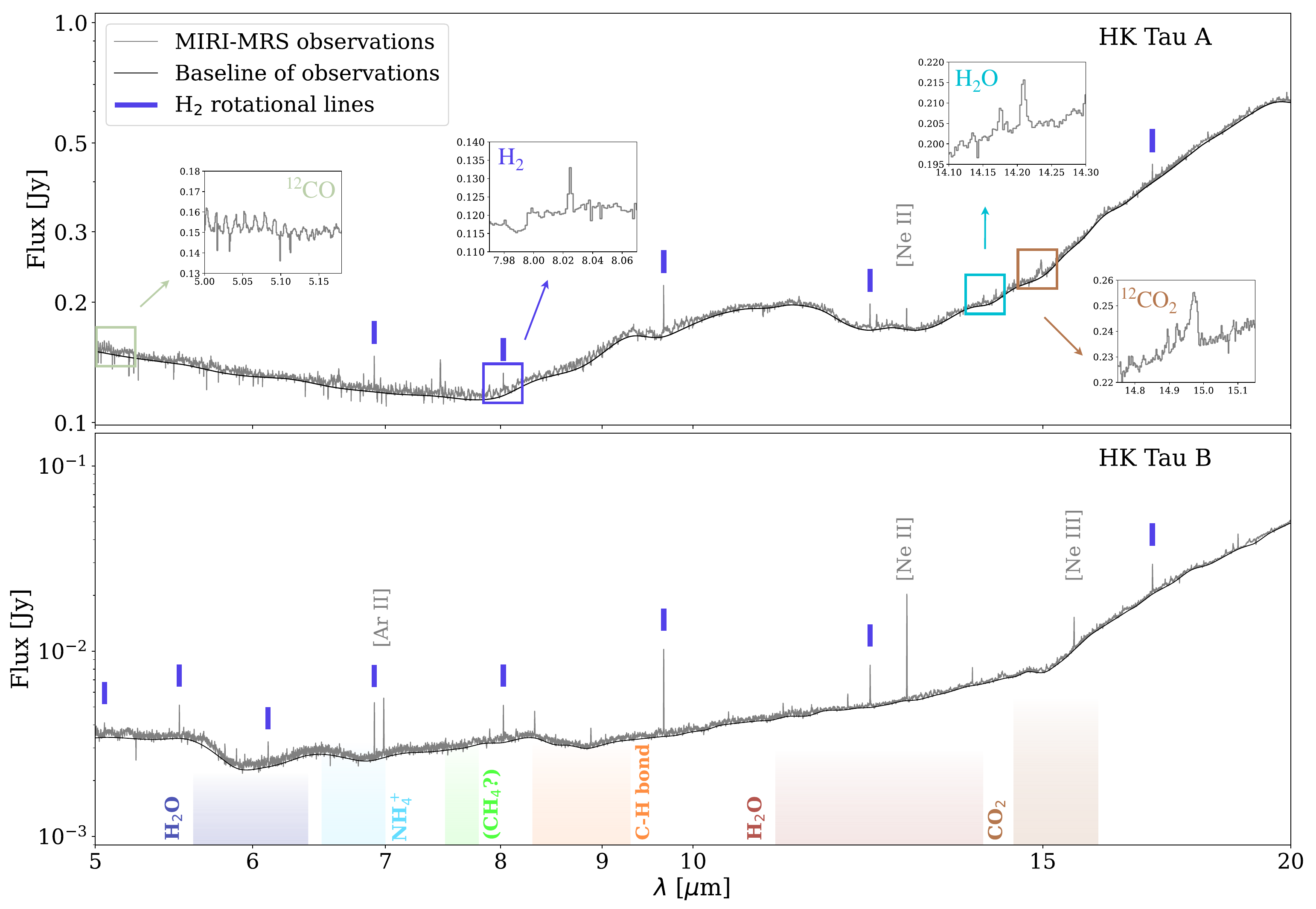}
    \caption{Integrated MIRI-MRS spectrum of HK Tau A (top) and B (bottom) in grey, with the black line representing the continuum baseline. In the top panel, the insets zoom into wavelength regions where molecular emission from \ch{^{12}CO}, \ch{H2O}, \ch{^{12}CO2}, and \ch{H2} is visible. In the bottom panel, the shaded bands in the  indicate the main ices absorption regions. \ch{H2} rotational lines are marked with blue ticks, and gas emission features are labelled where present.}
    \label{fig:spectrum}
\end{figure*}

The paper is structured as follows. In Section \ref{sec:observations} we describe the observations and data reduction process; in Section \ref{sec:spectra} we show the mid-infrared spectra of both sources and analyse their molecular and atomic emission lines, as well as the ice absorption features in the spectrum of B; in Section \ref{sec:ext_emission} we focus on the molecular hydrogen extended emission and derive a temperature and density 2D map; finally, we discuss the results in Section \ref{sec:discussion} and draw our conclusions in Section \ref{sec:conclusions}.

\section{Observations and data reduction}\label{sec:observations}

The HK Tau system was observed with JWST/MIRI-MRS on February 27th-28th, 2023, as part of the MINDS GTO Program (PID: 1282, PI: Th. Henning, \citealt{kamp2023chemical-67c, henning2024minds-d7e}). A four-point dither was performed in the positive direction. The total exposure time was 21.1 min per grating setting (for a total of 1.03 h). The data were reduced with the MINDS pipeline\footnote{The pipeline and its documentation are available at \texttt{https://github.com/VChristiaens/MINDS}} \citep{christiaens2024minds-9ba}, a hybrid pipeline that combines routines from the standard JWST pipeline (\citealt{bushouse2024jwst-80c}, v1.14.0) and the VIP package \citep{gomez-gonzalez2017vip-513, christiaens2023vip-679}. 

\begin{figure*}[htbp]
    \centering
    \includegraphics[width=0.95\linewidth]{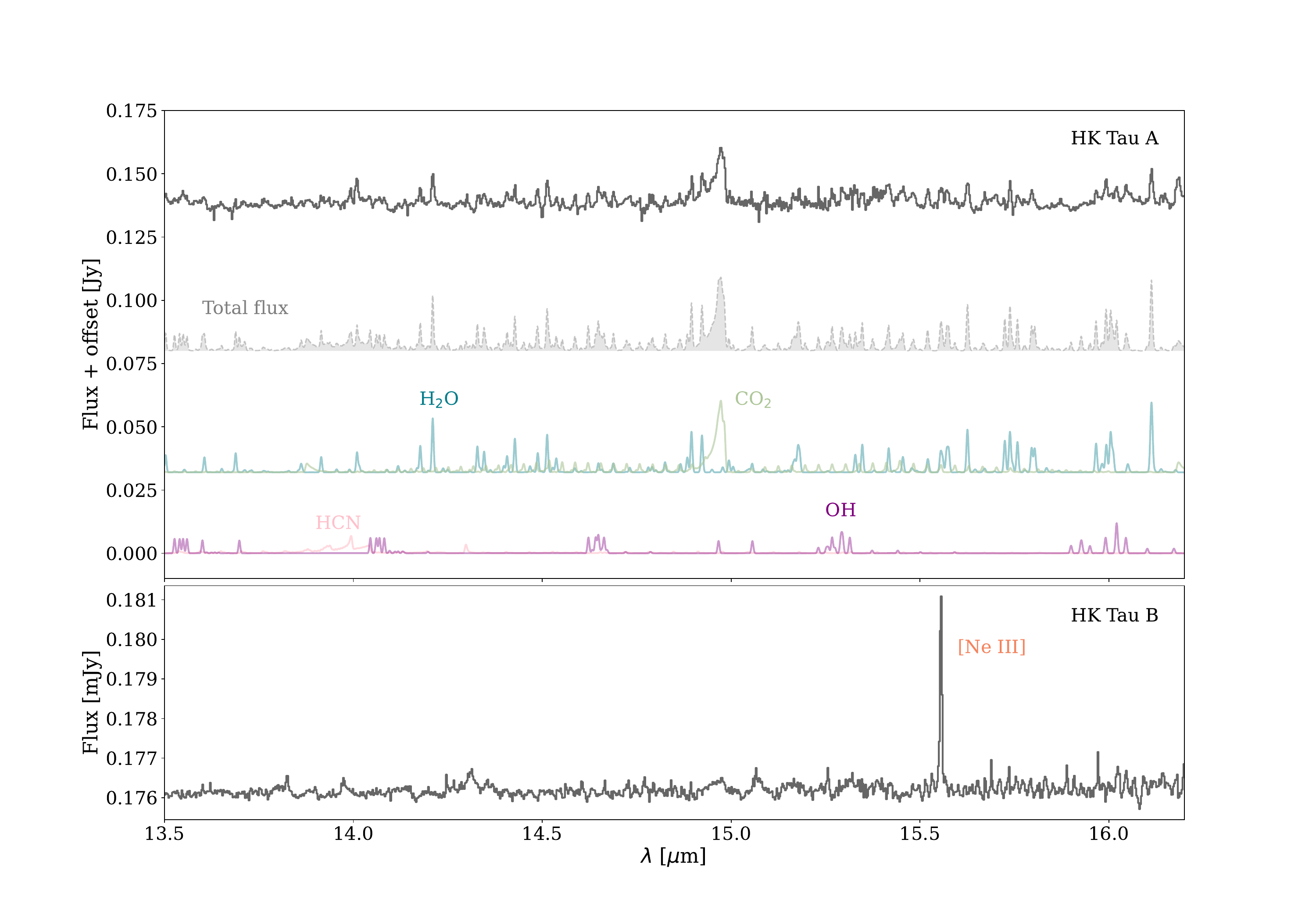}
    \caption{Comparison of the continuum-subtracted spectrum of HK Tau A (top panel, dark grey) and B (bottom panel) in the 13.5-16.2\micron\ range. The primary shows emission lines from several molecular components (\ch{H2O}, \ch{^{12}CO2}, \ch{HCN}, and \ch{OH}), to which we show the total fit (light grey) as well as the single molecular components (colored lines); on the other hand, in HK Tau B the noise level is too high to make any confident claim for molecular emission, and we only see the bright [Ne III] line.}
    \label{fig:radexpy_A}
\end{figure*}

The separation between the sources in the HK Tau system is \ang[]{;;2.4}, which does allow to spatially resolve them at the wavelengths of MIRI-MRS; however, at longer wavelengths the point-spread function (PSF) becomes wider, and the wings of the PSF of the sources overlap with each other, blending their fluxes. To extract the spectrum of HK Tau A and B, we apply the methodology of \cite{kurtovic2026minds-5c9}, which combines forward modelling with a theoretical PSF for each channel and aperture photometry to disentangle the emission of each of the two components. The emission of a source can extend further than the FWHM of the PSF, and the wings of the PSF of each source can overlap with the location of the other. Thus, the main reason for subtracting a forward model is to avoid contaminating the spectrum of a source with emission from the PSF wing of its companion. The separation between the sources is computed by fitting their position at each wavelength in band 1 short, and the centroid of the PSF and background level are free parameters, determined with a Markov Chain Monte Carlo (MCMC) approach. The highest amplitude residuals are found within one FWHM of the PSF, which is enclosed in the aperture for each disc; therefore, these residuals are included in the spectra. We refer to Appendix A of \cite{kurtovic2026minds-5c9} for a detailed description of the method. We estimate the aperture size as $\theta = 1.22$ $\lambda/D$, where $\lambda$ is the channel wavelength and $D$ is the diameter of JWST; as in \cite{kurtovic2026minds-5c9}, we use $2\theta$ for the primary and $\theta$ for the secondary. We confirm that these apertures are large enough to include all relevant residuals. At the longest wavelength of interest (20 \micron), the aperture is 3 times smaller than the binary separation. Assuming that all the noise is stochastic, in the line-free emission range from 16.34 to 16.36\micron\ we find a noise level of 2 mJy (relative to the combined spectra); however, because of the unconventional method to recover the spectra, our uncertainty is dominated by the systematic uncertainties of the extraction method rather than the sensitivity of the observations. The continuum emission in the 1D spectrum is estimated iteratively with a Savitzky-Golay filter by fitting a third-order polynomial, masking spikes deviating by $2 \sigma$ in the positive direction and $3 \sigma$ in the negative direction of the spectra \citep{temmink2024minds-3d3}. The baseline of the filtered spectrum is then determined using \texttt{PyBaselines} \citep{erb2022pybaselines-b34} and sdither as background subtraction method.

\section{Mid-infrared spectra}\label{sec:spectra}

Figure \ref{fig:spectrum} shows the MIRI-MRS spectrum of HK Tau A (top panel) and B (bottom panel). The secondary is dimmer, with a flux lower than the primary by a factor varying between 50 and 10 (at 5 and 20\micron\ respectively). Both sources show molecular hydrogen emission lines; while only the S(1) to S(5) transitions, out of the eight that fall in the MIRI range, are visible in the spectrum of A, we can see all of them in B (although with different intensities). Furthermore, at first glance we see strong atomic ion emission in B from both [Ar II] at 6.98\micron, [Ne II] at 12.81\micron, and [Ne III] at 15.55\micron; in A on the other hand, only [Ne II] is immediately visible and far less bright relative to the continuum. Contrary to all five T Tauri edge-on discs observed with MIRI-MRS so far, we do not see any signature of Polycyclic Aromatic Hydrocarbon (PAH) emission (expected to peak at 6.2, 7.7, and 11.3\micron; see \citealt{dartois2025edge-on-a02} for a template) in either source. The edge-on configuration of B allows ices to show in the spectrum as absorption bands, which we mark with coloured shaded areas. In the following, we dive deeper in the characterisation and discussion of the spectral features, from the molecular (Section \ref{subsec:gas_lines}) and the atomic ions emission lines (Section \ref{subsec:ionised_atoms}), to the ice features in the secondary (Section \ref{subsec:spectrum_ices}).

\begin{figure*}
    \centering
    \includegraphics[width=0.95\textwidth]{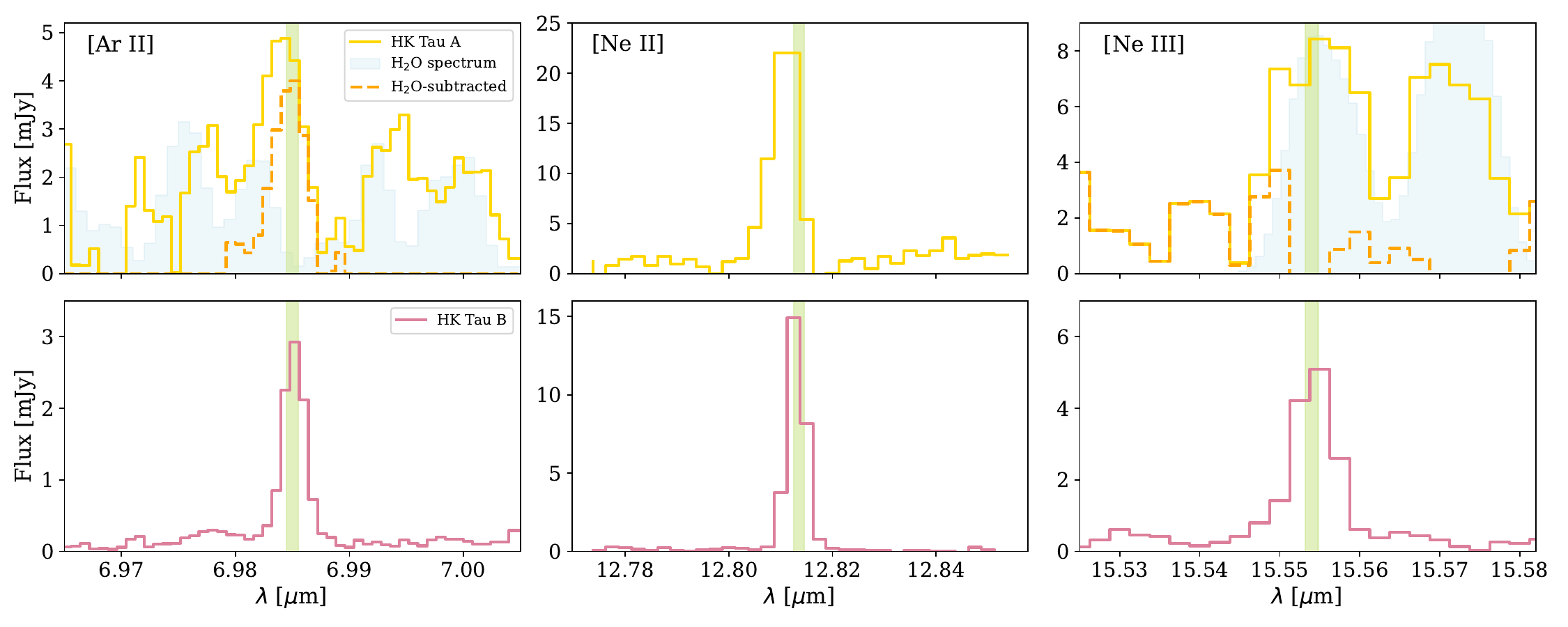}
    \caption{Emission at the wavelength of [Ar II], [Ne II], and [Ne III] from HK Tau A (top panel, yellow line) and B (bottom panel, pink line), from left to right. For [Ar II] and [Ne III] the top panel includes the fitted water spectrum for HK Tau A, as the lines fall in a water-rich wavelength region; we also show the resulting water-subtracted spectra for HK Tau A (orange, dashed line). The rest frame wavelength is highlighted in green.}
    \label{fig:forbidden_lines}
\end{figure*}

\subsection{Gas lines}\label{subsec:gas_lines}

Figure \ref{fig:radexpy_A} shows the continuum-subtracted spectrum of both HK Tau A (top panel, dark grey) and B (bottom panel) in the 13.5 - 16.2\micron\ range. We select this wavelength range due to its richness in molecular features \citep{carr2008organic-95f}. While the total spectrum of B is richer in forbidden line emission and ice absorption bands, it is extremely poor in high contrast molecular lines - such as those associated with the warm water emission of other T-Tauri discs (e.g., \citealt{gasman2023minds-c35, grant2024minds-a9e, temmink2024minds-3d3, banzatti2025water-438}) - in line with the results for other binary systems (\citealt{kurtovic2026minds-5c9}; see Section \ref{subsec:discussion_molecules}). The continuum-subtracted spectrum of B does show some wiggles, including a bump at $\sim$ 15\micron\ compatible with \ch{^{12}CO2}; however, the noise level is too high to be able to make a confident claim on molecular emission. The only convincing feature we see in the secondary spectrum at this wavelength range is the [Ne III] line at 15.5\micron. On the other hand, the primary shows several gas emission lines: it is particularly straightforward to identify the strong \ch{^{12}CO2} Q-branch feature at $\sim$ 15\micron\ and a few water lines.

To constrain the molecular composition of HK Tau A, we fit the continuum-subtracted spectrum with Local Thermodynamical Equilibrium (LTE) 0D slab models \citep{grant2023minds-eea, perotti2023water-6d2, tabone2023rich-ab1}. The 13.5-16.2\micron\ region, falling in band 3 medium and long, allows us to study the emission of \ch{H2O}, \ch{CO2}, \ch{HCN}, and \ch{OH}; we set the spectral resolution of the slab models in accordance with the most recent estimations (2500 - \citealt{pontoppidan2024high-contrast-766, banzatti2025water-438}; see also the original commissioning estimates by \citealt{argyriou2023jwst-f58}) and use the line transitions derived from the HITRAN database \citep{gordon2022hitran2020-7b5}. We use as line width $\Delta V =$ \SI{4.71}{\kmpers} as in \cite{salyk2008h2o-305, salyk2011co-094}. Following \cite{grant2023minds-eea}, we perform the fit by chi-squared minimisation iteratively, subtracting the contribution of each molecule after fitting for it choosing common bright lines, in the order \ch{H2O}, \ch{OH}, \ch{CO2}, and \ch{HCN}; we detect emission from all of them, and show the fitted spectrum in the top panel of Figure \ref{fig:radexpy_A}. Each slab model has three free parameters that are varied to fit the emission features: the temperature $T$, the column density $N$, and the emitting area $\pi {R_{slab}}^2$ which is characterised by an emitting radius $R_{slab}$. The minimum allowed temperature was 100 K, while the maximum emitting radius was 10 au. The results of the fit are reported in Table \ref{tab:appendix_fit}: we find \ch{H2O} at 725 K with a column density of $2.15 \times 10^{18} \rm{cm}^{-2}$, while \ch{CO2} has a temperature of 350 K and a column density of $4.64 \times 10^{17} \rm{cm}^{-2}$. Based on \textit{Spitzer} observations, \cite{bosman2017co2-b31} identified HK Tau as a \ch{CO2}-only source (i.e., without water signatures); the \textit{Spitzer} observations did not have high enough sensitivity and spatial resolution to separate the two components of the binary system\footnote{For a comparison of the \textit{Spitzer} IRS and JWST/MIRI-MRS spectra, see Appendix \ref{appendix:comparison_Spitzer}.}, while with JWST/MIRI we can now see that the bulk of the \ch{CO2} emission comes from the primary. Furthermore, HK Tau A is rich in molecular emission in general, and shows water features as well. The \ch{^{12}CO2} Q-branch stands out in the spectrum compared to the water features, in line with the empirical definition of \ch{CO2}-dominated sources; however, despite the strong \ch{^{12}CO2} emission, we do not detect any \ch{^{13}CO2} nor other minor isotopologues - as opposed to the case of MY Lup \citep{salyk2025emission-ba3}.

\subsection{Forbidden emission lines from atomic ions}\label{subsec:ionised_atoms}

Figure \ref{fig:forbidden_lines} shows the continuum-subtracted spectrum of HK Tau A (yellow, dashed line) and HK Tau B (pink, solid line) in wavelength ranges around at the centre of the [Ar II], [Ne II], and [Ne III] forbidden emission lines. These lines trace ionised outflowing gas and are interpreted as signatures of disc winds and jets, depending on their velocity and morphology; in particular, [Ne II] is traditionally classified into high or low- velocity component (HVC or LVC, associated to jets and winds respectively, \citealt{pascucci2020evolution-752}), depending on whether the line centroid is shifted by more or less than \SI{30}{\kmpers}.

The lines in HK Tau A are more noisy and contaminated, mostly by water lines, due to the rich molecular emission; as [Ar II] and [Ne III] fall in water-rich wavelength ranges, Figure \ref{fig:forbidden_lines} also shows the best-fit water spectrum (light blue, shaded) as well as the water-subtracted HK Tau A spectrum (orange, dotted line). After the water lines subtraction, HK Tau A preserves emission associated to [Ar II], while the emission at the rest frame wavelength of [Ne III] is entirely removed. Therefore, we conclude that there is no [Ne III] emission coming from the primary disc.

\begin{table}[h]
    \centering
    \caption{Forbidden atomic lines fluxes and velocities in HK Tau A and B.}
    \begin{tabular}{c c c c c}
    
        \hline

        \rule{0pt}{2.5ex}\multirow{2}{*}{Line} & $\lambda$ & Source & Flux & $v$ \\
         & [\textmu m] & & [$10^{-15}$ erg cm$^{-2}$ s$^{-1}$] & [km s$^{-1}$] \\[1ex]
        
        \hline
        \rule{0pt}{2ex}\multirow{2}{*}{[Ar II]} & \multirow{2}{*}{6.98} & A & 0.9 $\pm 10 \%$ & 5 $\pm$ 8 \\
         & & B & 0.4 $\pm 10 \%$ & 12 $\pm$ 15 \\
         \hline
        \rule{0pt}{2ex}\multirow{2}{*}{[Ne II]} & \multirow{2}{*}{12.81} & A & 3 $\pm 10 \%$ & -35 $\pm$ 10 \\
         & & B & 1.2 $\pm 10 \%$ & $\geq$ 16 $\pm$ 10 \\
         \hline
        \rule{0pt}{2ex}\multirow{2}{*}{[Ne III]} & \multirow{2}{*}{15.55} & A &  // &  // \\
         & & B & 0.4 $\pm 10 \%$ & 8 $\pm$ 30 \\

    \end{tabular}
    \tablefoot{// indicates non detections. The statistical uncertainties on the line fluxes is unrealistically small, so we take 10$\%$ of the measured flux as a conservative estimate. The tabulated velocity uncertainties are at 3$\sigma$.}
    \label{tab:line_fluxes}
\end{table}

To study the resulting forbidden line emission, we fit a Gaussian profile to each line and determine their total flux and velocity relative to the rest wavelength; we report the results in Table \ref{tab:line_fluxes}. In the primary, we find redshifted [Ar II] and blueshifted [Ne II]; in the secondary instead, we see redshifted emission for all three lines, consistent with having a radial velocity close to the rest velocity. The [Ne II] in HK Tau B shows a LVC wind with \SI{16.3}{\kmpers}, while HK Tau A has a HVC with \SI{-35}{\kmpers} instead. Note that the high inclination of HK Tau B is such that, in the case of emission from a collimated flow, the measured velocity represents the projected velocity, thus requiring a correction factor of $1/\cos{(i)}$. This translates to a factor 8 for the inclination of this source (corresponding to a velocity of $\sim$\SI{130}{\kmpers}), which would lead this to be classified as an HVC jet, rather than an LVC wind (see \citealt{pascucci2020evolution-752} for examples of HVC sources that look like LVC in projection). However, if the emission comes from a less collimated flow (which can be determined from its morphology), then the correction factor may be less, and the flow may truly be an LVC. We discuss this further in Section \ref{subsec:discussion_wind}.

\subsection{Ice features in HK Tau B}\label{subsec:spectrum_ices}

The strongest ice absorption feature detected in the spectrum of HK Tau B is the bending mode of water at 6.2\micron; water ice is also identified in the libration mode at 13.6\micron, although this band is significantly weaker. The second most evident absorption feature is \ch{CO2} at 15.2\micron, while the broad dip around 6.85\micron\ is usually associated to \ch{NH4+}. Finally, between 8.3 and 9\micron, we see an absorption band which includes two separate dips at 8.50 and 8.85\micron\ respectively. In the following, we discuss the details and implications of each of these signatures.

\begin{enumerate}

    \item Water ice. In our spectrum, the bending mode of water is stronger than the libration mode; as pure water ices are expected to show the opposite behaviour, with a stronger libration than bending mode \citep{berg2007effects-ea3}, the 6.2\micron\ absorption feature is likely a blend of \ch{H2O} and \ch{CH3OH}.

    \item \ch{NH4+}. The broad 6.85\micron\ band likely includes a contribution by \ch{CH3OH}, with its \ch{C-H} deformation mode. This feature, however, could also be due to \ch{HCOOH}, \ch{H2O}, or \ch{NH3}, together with the \ch{NH4+} bend \citep{keane2001ice-fe4, schutte199660-e1e, slavicinska2025ammonium-cf2}. The production of \ch{NH4+} from the dissociation of \ch{NH4CN} is associated to \ch{OCN-} absorption at around 4.6\micron\ \citep{gerakines2024sublimation-59c}, which we indeed see in the NIRSpec spectrum of HK Tau B (Smith et al. in prep.).

    \item \ch{CH4}. Contrary to other sources (such as HH 48 NE, \citealt{sturm2023jwst-545}) we do not confidently see the \ch{CH4} feature at 7.71\micron; we highlight the expected location in the spectrum, where we have a barely visible dip.

    \item Silicate feature/C-H bend. The absorption band at 8.3-9\micron\ is consistent with the silicate stretching band that usually peaks at 9.7\micron\ in systems at lower inclinations (\citealt{henning2010cosmic-c52}, visible in HK Tau A in emission with the characteristic shape). This shift to shorter wavelengths was observed also in HH 48 NE \citep{sturm2024jwstmiri-3fe}, and predicted by \cite{sturm2023-modelingices} with radiative transfer models: the offset is due to the presence of an extra emission component coming from the inner disc, which is scattered in a similar way to the continuum. We note that inclination has been shown to impact the shape and the wavelength of the water ice bands as well \citep{martinien2025role-fab}. It is nonetheless tempting to try and attribute the two distinct peaks we see in HK Tau B to specific ice features: their weakness, however, makes it difficult to associate them with anything beyond the general \ch{C-H} bond bending mode around 8.6\micron.

\end{enumerate}

\section{Extended \ch{H2} emission}\label{sec:ext_emission}

\begin{figure*}
    \centering
    \includegraphics[width=0.99\textwidth]{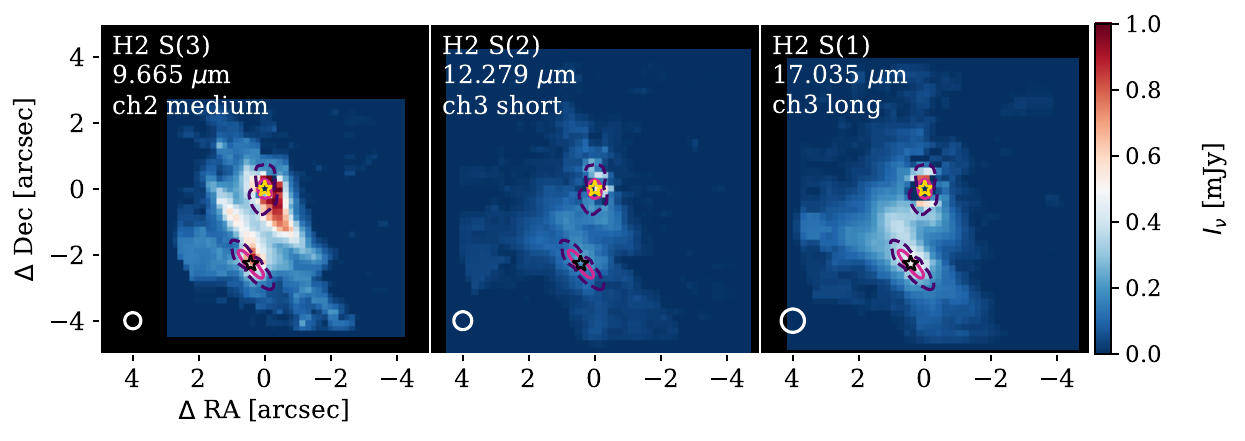}
    \caption{Moment 0 maps of the S(1), S(2), and S(3) \ch{H2} extended lines. The yellow and black stars mark the coordinates of the centre of HK Tau A and B respectively. The white circles in the bottom left corner of each panel show the full width at half maximum of the PSF at the corresponding band. North is up, east is to the left. The pink solid and purple dashed contours respectively represent the extent of the ALMA continuum emission at 0.87 mm (published in \citealt{villenave2020observations-5b8}) and the peak emission map of CO J=2-1 (published in \citealt{rota2022observational-f65}; estimated with the quadratic method of \texttt{bettermoments} \citep{TeagueBetterMoments2018, TeagueBettermoments2_2018}).}
    \label{fig:mom0_threechannels}
\end{figure*}

We detect extended emission for the \ch{H2} transitions from S(1) to S(5) around both sources, as well as the other three falling in the MIRI-MRS range - S(6) to S(8)) around HK Tau B. Figure \ref{fig:mom0_threechannels} presents the moment 0 maps at the three longest wavelengths - corresponding to S(3), S(2) and S(1) from left to right, falling in band 2 medium (9.665\micron), 3 short (12.279\micron), and 3 long (17.035\micron) respectively. For the moment maps of all \ch{H2} transitions, see Appendix \ref{appendix:RD_and_maps}.

The most prominent feature of the extended emission in Figure \ref{fig:mom0_threechannels} is the 'X'-shaped structure around HK Tau B: the location and inclination of the disc are compatible with this emission originating from the surface of the disc around HK Tau B, therefore tracing the disc surface itself or a wind. We computed the semi-opening angle by rotating the S(1) moment 0 map and determining the maximum brightness at \SI{1.5}{\arcsecond} from HK Tau B - the largest distance where we can find maxima across both sides of the emission. We find a semi-opening angle of \ang{77} $\pm$ \ang{13} for the north-west surface and \ang{69} $\pm$ \ang{12} for the south-east one. The higher brightness of the north-east filament (most easily seen in S(1) and S(3) but present at all wavelengths) is likely due to the three-dimensional position of the sources: as noted by \cite{koresko1998circumstellar-281}, and visible already in the scattered light image presented in \cite{mccabe2011spatially-5ef}, the disc around HK Tau B is not symmetric and the north side is brighter than the south side, which is compatible with a higher irradiation coming from HK Tau A. The brightest spot visible in scattered light is on the northern-eastern side of the disc, at the same location as the brightest emission in the \ch{H2} map. To investigate the nature of the 'X'-shaped \ch{H2} emission, we compare its extent with that of the ALMA continuum and \ch{CO}, represented in Figure \ref{fig:mom0_threechannels} by the pink solid and purple dashed contours respectively. As expected in edge-on discs, the continuum emission is concentrated in a horizontally thin structure, tracing the dust-rich midplane; the keplerian CO is slightly more extended, both vertically and radially, but still less than the \ch{H2}. Furthermore, we do not see signatures of the 'X'-shape, pointing in the direction of the \ch{H2} being entrained in a wind rather than sitting at the disc surface. To constrain the temperature $T$ and column density $N$ of the \ch{H2}, we perform a rotational diagram analysis (see Section \ref{subsec:analysis_rot_diagram}) at each pixel of the moment 0 maps. As the \ch{H2} transitions span from channel 1 short to channel 3 long (subbands 1A to 3C), the resulting maps have different pixel sizes and angular resolution: for this reason, we perform a PSF matching and resampling algorithm before the fit. 

\begin{figure*}
    \centering
    \includegraphics[width=\textwidth]{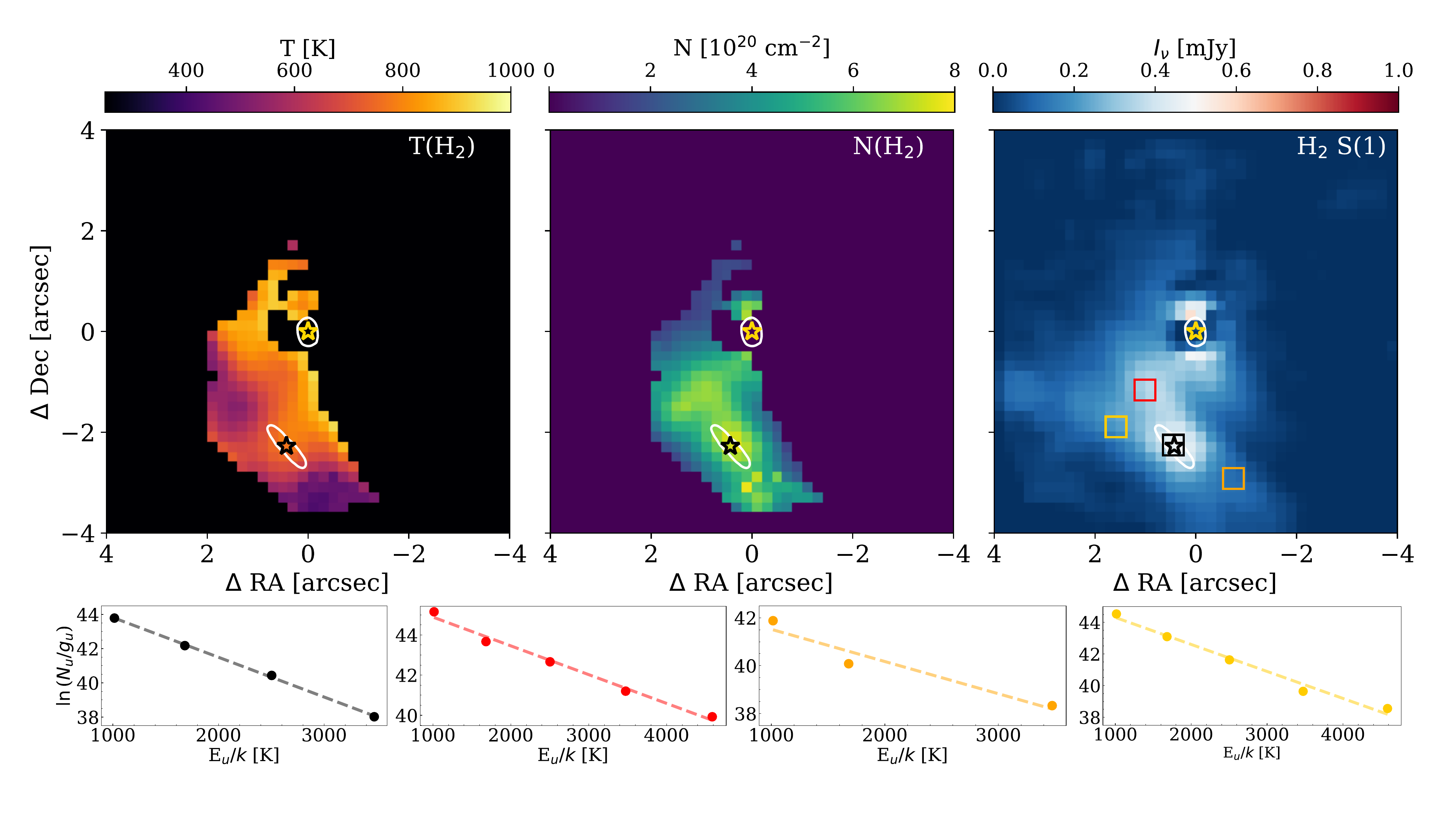}
    \caption{Pixel-by-pixel temperature (left) and total column density (centre) maps of the extended \ch{H2} emission in the surroundings of HK Tau B, compared to the S(1) moment 0 map (right). The maps are centred at the location of HK Tau A, marked by the yellow star, while the black star represents the location of HK Tau B. The white contours represent the extent of the ALMA continuum emission as in Figure \ref{fig:mom0_threechannels}. The bottom row shows the rotational diagram at each of the four locations marked in the S(1) moment 0 map by the coloured squares.}
    \label{fig:analysis_T_N_maps}
\end{figure*}

\subsection{PSF matching and resampling}\label{subsubsec:analysis_PSFmatch_resample}

The PSF of MIRI-MRS is larger at longer wavelengths. As the \ch{H2} rotational lines span from 5.053 to 17.035\micron, the maps have different PSFs (and therefore different resolutions) that need to be matched before performing any spatial analysis. As we are interested in the extended emission between the two sources, and therefore do not need a particularly high spatial resolution, we decided to match the lowest one (corresponding to the S(1) line). This also prevents the introduction of sampling artifacts. We downgrade the resolution of all lines to that of S(1) by convolving the maps with an appropriate (i.e., FWHM-based), wavelength-dependent Gaussian kernel. Once all maps are at the same resolution, we ensure a homogeneous sampling by interpolating them on a same grid with even spacing. We chose to have four pixels per beam, leading to a sampling of 200 mas per pixel. 

\subsection{\ch{H2} rotational diagram}\label{subsec:analysis_rot_diagram}

To obtain an estimate of the gas temperature and column density, we perform a rotational diagram analysis \citep{goldsmith1999population-437} on the \ch{H2} rotational lines. Such analysis can be applied to populations of molecules under the assumption of (1) local thermodynamic equilibrium (LTE), (2) optically thin emission, and (3) that the column of \ch{H2} contributing to the emission has a single $T$ and $N$ at all wavelengths. In LTE (1), the column density at the upper state excitation level $N_u$ can be written as

\begin{equation}
    N_u = \frac{N_{\mathrm{tot}}}{Q(T)} g_u \exp{\left( -\frac{E_u}{k T} \right)},
    \label{eq:analysis_RD1}
\end{equation}

\noindent where $N$ is the total column density, $Q(T)$ is the partition function at the excitation temperature $T$, $E_u$ upper state with energy with statistical weight $g_u$, and $k$ is the Boltzmann constant. Furthermore, if the emission is optically thin (2), the upper state column density is directly linked to the observed flux $F_u$,

\begin{equation}
    N_u = \frac{4 \pi F_u}{A_u h \nu \Omega},
    \label{eq:analysis_RD2}
\end{equation}

\noindent where $\Omega$ is the emitting area and $A_u$ is the Einstein rate coefficient for spontaneous emission. Substituting $N_u$ in Equation \ref{eq:analysis_RD2}, one can then write a linear relation (in logarithmic space) between $F_u$ and $E_u/k$,

\begin{equation}
    \ln{\left( \frac{4 \pi F_u}{A_u h \nu \Omega} \frac{1}{g_u} \right)} = - \frac{1}{T} \frac{E_u}{k T} + \ln{\left( \frac{N_{\mathrm{tot}}}{Q(T)} \right)}
    \label{eq:analysis_RD_fit}
\end{equation}

\noindent with a slope $m = - 1/T$ and an intercept $q = \ln(N_{\mathrm{tot}}/Q(T))$. The transitional constants $E_u$, $g_u$, and $A_u$ used in the rotational diagram analysis for the \ch{H2} lines are given in Table \ref{tab:analysis_rotvi_constants}; the (normal) partition function $Q(T)$ is taken from \cite{popovas2016partition-a55}\footnote{Note that \cite{popovas2016partition-a55} define the ortho-para ratio as 3/4:1/4 rather than 3:1, therefore the resulting partition function needs to be normalised by a factor 4.}.

\subsection{Temperature and column density 2D maps}\label{subsubsec:analysis_T_N_maps}

Once all maps have the same resolution, as well as number of pixels and pixel sampling, we perform a one-component rotational diagram analysis in each pixel for all transitions to obtain a temperature and column density value. The top row of Figure \ref{fig:analysis_T_N_maps} shows the resulting 2D $T$ (left) and $N$ (centre) maps, next to the S(1) emission for ease of comparison (right); in the bottom row, we report the result of the fit for four specific locations (marked in the moment 0 map by coloured squares) to validate the assumption of one-component fits. The maps are centred on the location of HK Tau A, marked by the yellow star, while the centre of HK Tau B is represented by the black star. The white contours show the extent of the ALMA continuum emission. We applied a mask on the line intensity to remove all pixels with a flux lower than \SI{0.1}{\mJy}, and we discarded all pixels with less than three \ch{H2} lines detected to ensure a more reliable fit.

The region around HK Tau B where the continuum emission originates shows a temperature of around \SI{800}{\K}, which extends to the 'X' structure visible in the S(1) line centred in B. Parallel to the midplane of the disc around B we see two lobes with a lower temperature of around \SI{500}{\K}. This lower temperature could be a result of the geometry of the system, with the disc obscuring its surrounding exactly beyond the major axis of the disc. The column density in the higher temperature region is around $2 \times 10^{21}$ cm$^{-2}$, and in general across the system it varies between $\sim 0.5-2 \times 10^{21}$ cm$^{-2}$. The temperature is in line with other Class II sources showing extended \ch{H2} emission, varying between $\sim$ 400 and $\sim$ 1100 K \citep{narang2026characterizing-83c}; the column density, on the other hand, is somewhat larger than the maximum value of in the sample of \cite{narang2026characterizing-83c} ($\log_{10}(N) = 19.5$).

\section{Discussion}\label{sec:discussion}

\subsection{The impact of disc inclination on molecular emission}\label{subsec:discussion_molecules}

\begin{figure*}
    \centering
    \includegraphics[width=0.95\linewidth]{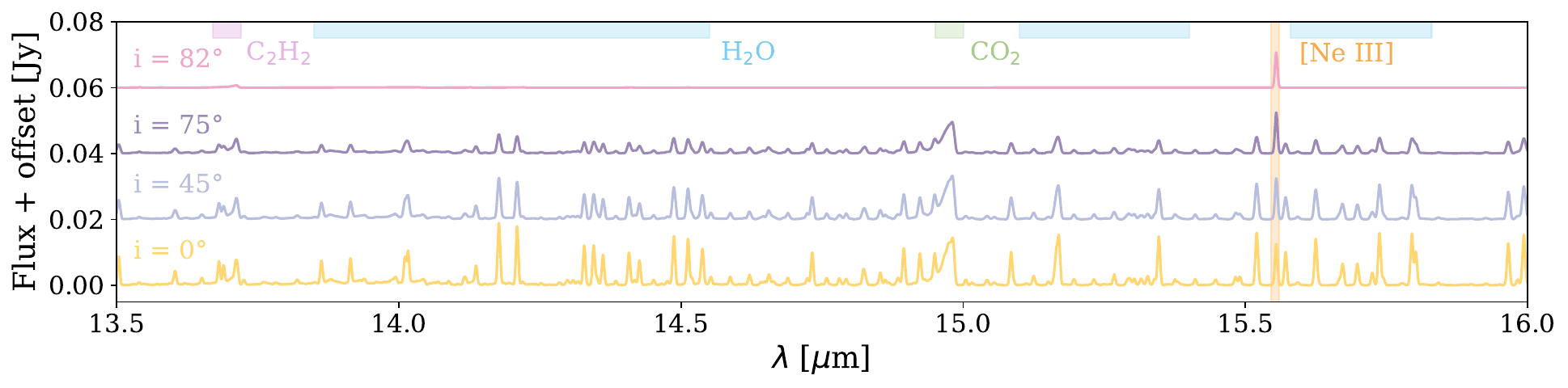}
    \caption{Synthetic spectra of a fiducial T Tauri disc model \citep{arabhavi2026molecular-927} showing the impact of disc inclination on the molecular features at 13.5-16\micron.}
    \label{fig:RTmodels}
\end{figure*}

The dichotomy in molecular features of the two sources, with a line-rich primary and line-poor (except for extended emission) secondary, is similar to the trend found by \cite{kurtovic2026minds-5c9} for the three binary systems VW Cha, WX Cha, and RW Aur, as well as the suggested case of DF Tau \citep{grant2024minds-a9e}. In all of these systems, the molecular emission at MIRI-MRS wavelengths is dominated by the primary star, which is on average more than a factor 7 brighter in flux with respect to the secondary. This evidence has been interpreted as a consequence of secular evolution, with tidal interactions truncating the discs - more effectively around the secondaries - and therefore draining the supply of molecules to the inner disc, where they then emit at MIRI-MRS wavelengths.

HK Tau shares the line-rich primary and line-poor secondary behaviour, but fundamentally differs from the systems mentioned above in that (i) the mass ratio is close to one and (ii) the secondary is edge-on. This has two main implications: on one hand, with the mass ratio being close to unity, the distinction between primary and secondary is more blurry - and ultimately determined by the inclination, which influences the luminosity of the two sources; the high inclination is also likely to have an impact on the luminosity of the lines originating in the inner disc, as they have to go through multiple scatterings in the thick disc midplane.

To assess the impact of disc inclination on the expected molecular emission, we have produced synthetic MIR spectra from a fiducial T Tauri disc model (published in \citealt{arabhavi2026molecular-927} and computed with \textsc{ProDiMo}, \citealt{woitke2009prodimo}), running radiative transfer models with Fast Line Tracers (FLiTs, \citealt{woitke2018modelling-0fb}) to calculate the spectra at different viewing angles. Figure \ref{fig:RTmodels} shows the expected spectral features in the 13.5-16\micron\ wavelength range for inclinations increasing from \ang{0} to \ang{82} (corresponding to the inclination of HK Tau B), without accounting for scattering effects. The molecular flux is reduced at increasing inclination, but remains detectable up to $\sim$ \ang{75}; at nearly edge-on configurations (> \ang{75}), instead, the molecular lines disappear from the spectrum, which only shows signatures of ionised atomic emission. This happens because the atomic ion emission originates in the upper layers of the disc -- either the disc surface itself or winds and outflows -- as opposed to the molecular component, which comes from the midplane instead; as a consequence, the molecular emission is hidden behind a larger amount of material compared to the atomic ions, which remain visible (albeit damped) even in the edge-on configuration. Interestingly, the inclination threshold to maintain molecular emission coincides with the sweet spot for the detection of ice absorption features: this means that only sources at the interface between the face-on and edge-on configuration may show both gas- and ice-phase features.

It is worth noting that other edge-on sources in the literature do show scattered molecular emission. Nonetheless, their spectra remain less rich in that only CO emission is present \citep{arulanantham2024jwst-ecf, sturm2024jwstmiri-3fe, bergner2026jwst-00e}, with the exception of Tau 042021 where both CO and water are detected \citep{arulanantham2024jwst-ecf}. This suggests that in HK Tau B, where no molecular emission is present at all, a combination of the high inclination and other geometrical effects (e.g., misaligned inner disc or substructures) results in reduced scattering. In fact, the line-poor spectrum of RW Aur B \citep{kurtovic2026minds-5c9} has been interpreted as a consequence of the presence of a cavity in the inner disc: this hypothesis may hold for HK Tau B as well, supported by a marginal rise at the edges of the ALMA radial brightness profile \citep{villenave2020observations-5b8}.

The HK Tau system provides an unprecedented opportunity: the very close spectral type of the two sources combined with the complementary inclinations and reduced scattering effects allow to have a simultaneous view of the gaseous and solid components of the disc, under the assumption that the two sources underwent a similar evolution. To probe the chemical similarity between HK Tau A and B, we can exploit the edge-on nature of B to look for the icy counterpart of the molecules emitting in A. This is however non-trivial: while we do see \ch{H2O} and \ch{CO2} ices at 6, 13.6, and 15\micron, detecting signatures of \ch{OH} and \ch{HCN} ice is less straightforward. \ch{OH} is a minor component in solid phases, emitting a weak signal; the presence of \ch{HCN} instead is linked to \ch{NH4+} at 6.964\micron, which we do see in absorption (cfr. Figure \ref{fig:spectrum}), and \ch{OCN-} at 4.6\micron\ \citep{lacy198446-b63}. Despite these limitations, the gas- and ice-phase molecular signatures in the two sources point towards a similar chemical composition.

\begin{figure*}
    \sidecaption
    \includegraphics[width=12cm]{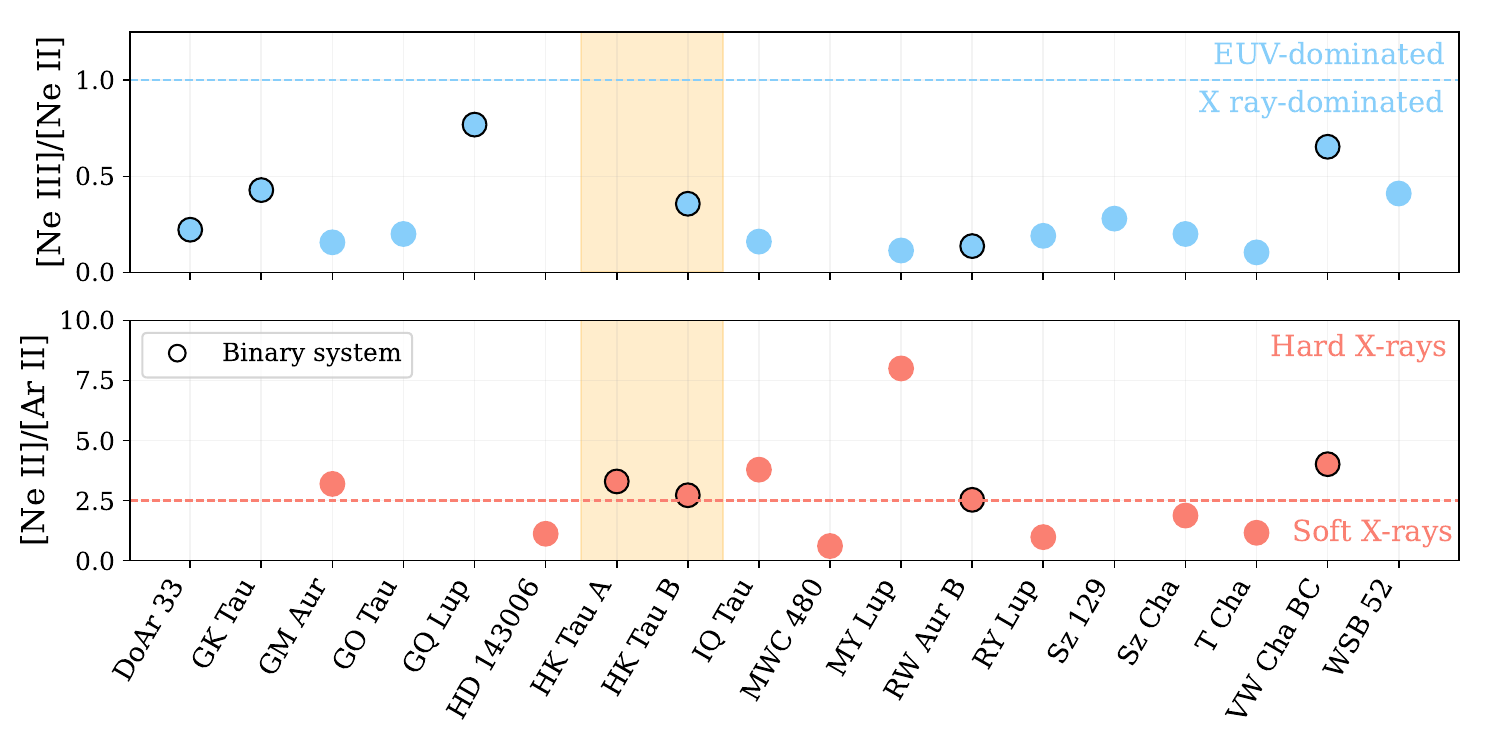}
    \caption{Comparison of the [Ne III]/[Ne II] (top panel) and [Ne II]/[Ar II] (bottom panel) line ratios for HK Tau A and B (orange shaded region) and the other published detections. Binary systems are marked with a black contour. The coloured dashed lines show the line ratio thresholds to determine the dominant irradiation source. Sz Cha from \cite{espaillat2023jwst-0e9}, T Cha from \cite{bajaj2024jwst-6d4}, RW Aur B and VW Cha BC from \cite{kurtovic2026minds-5c9}, all others from \cite{arulanantham2025jdisc-9a4}.}
    \label{fig:lineratios}
\end{figure*}

\subsection{Other spectral features: \ch{H2} and forbidden lines}\label{sec:discuss_other_sp_features}

The spectra of HK Tau A and B differ not only in the molecular lines, but also in the extended \ch{H2} and forbidden line emission. \ch{H2} is visible in the spectra of both stars, although the relative intensity of the lines is between a factor 5 and 10 larger in the primary. The spatial extension of the \ch{H2} emission is also different: in the moment 0 maps, it appears far brighter and more extended around B, and indeed, it traces not only the immediate surroundings of the central star, but also a radial extend $\sim 3$ times that of the ALMA continuum. The shape of the emission itself is expected to be due to the geometry, as the 'X'-shape is a signature of the high inclination; however, if the \ch{H2} emission were similarly strong also around the primary, we would expect to still see it spread over a larger area in the moment 0 maps.

The spectra also diverge in the presence and intensity of the forbidden atomic lines. While in HK Tau B we see bright [Ne II], [Ne III], and [Ar II] emission, only [Ne II] and [Ar II] are present in HK Tau A. [Ne II] has been routinely detected in several sources since the \textit{Spitzer}/IRS surveys \citep{pascucci2007detection-166, lahuis2007c2d-e8b, ratzka2007high-931, najita2010spitzer-bd0, gdel2010origin-13f, espaillat2013tracing-742}, as well as later observations at higher spectral resolution which allowed distinguishing between wind and jet origin (see, e.g., \citealt{pascucci2020evolution-752}). On the other hand, [Ne III] is harder to detect and remained rare in the \textit{Spitzer} era \citep{lahuis2007c2d-e8b, najita2010spitzer-bd0, szulgyi2012observational-9a7, espaillat2013tracing-742}; its emitting wavelength of 15.55\micron\ overlaps with warm ($\sim$ \SI{600}{K}) water features, which makes it challenging to disentangle in water rich sources even with JWST. The case of HK Tau, with a water-rich primary and line-poor secondary, shows a good example of this issue - in a similar fashion to the binary sample of \cite{kurtovic2026minds-5c9}; we therefore add HK Tau B to the sample of line-poor, Class II sources with detected [Ne III] emission (alongside T Cha, \citealt{bajaj2024jwst-6d4}, VW Cha BC, and RW Aur B, \citealt{kurtovic2026minds-5c9}). As we have shown in Figure \ref{fig:RTmodels}, forbidden lines are expected to remain visible (although with potentially different fluxes) across all inclinations; as these lines trace winds and jets, we suggest that, if there is a wind coming from HK Tau A, it is too cold or dense to be ionized and brightly emitting; \cite{pascucci2020evolution-752, pascucci2023role-55b} suggest that the presence of [Ne II] in an HVC may be evidence for a dense inner wind blocking the X-rays from ionising the Ne in the extended wind, therefore rendering the LVC absent.

\subsection{Forbidden line ratios}\label{subsec:line_ratios}

The detection and flux measurement of [Ne III] is especially relevant in comparison with [Ne II] to study the high-energy radiation absorbed by the emitting material in the inner disc in the context of photoevaporation. Combined with [Ar II], the three atomic lines trace either high- or low-velocity winds, depending on the strength of the disc accretion rate (high-velocity for $\dot M > 10^{-8}$ M$_{\odot} \rm{yr}^{-1}$, \citealt{pascucci2020evolution-752}; low-velocity otherwise, \citealt{hollenbach2009diagnostic-0d7, szulgyi2012observational-9a7, espaillat2013tracing-742, sellek2024modeling-bbd}). The values of the [Ne III]/[Ne II] and [Ne II]/[Ar II] line ratios are predicted to depend on the dominant source of irradiation: from photoevaporative models, we know that it is easier to produce [Ne III] over [Ne II] with EUV spectra, while X-ray ionisation produces multiply charged ions that then rapidly recombine and exchange charge with hydrogen atoms, leading to a higher [Ne II] production \citep{glassgold2007neon-366}; as a consequence, we expect [Ne III]/[Ne II] < 1 if X-ray irradiation dominates the disc ionisation, while on the other hand [Ne III]/[Ne II] > 1 would be indicative of an EUV-driven wind \citep{hollenbach2009diagnostic-0d7, espaillat2013tracing-742, bajaj2024jwst-6d4}. [Ne II]/[Ar II], on the other hand, is expected to be < 2.5 for EUV or soft X-ray radiation fields and > 2.5 for hard X-rays \citep{espaillat2023jwst-0e9}.

Figure \ref{fig:lineratios} shows the [Ne III]/[Ne II] (top panel) and [Ne II]/[Ar II] (bottom panel) ratios for HK Tau A and B, highlighted with the orange shaded regions, compared to all other protoplanetary discs with published line fluxes in the literature\footnote{To ensure consistency in the derived line intensities, we applied our pipeline to the sample of \cite{arulanantham2025jdisc-9a4}.}. The horizontal dashed lines represent the [Ne III]/[Ne II] = 1 and [Ne II]/[Ar II] = 2.5 thresholds. Binary systems are marked with a black outline. With a [Ne III]/[Ne II] ratio of 0.36, HK Tau B falls well within the X-ray irradiation regime; this seems to be the most frequently found in discs, with only a couple (GQ Lup and VW Cha BC, \citealt{arulanantham2025jdisc-9a4} and \citealt{kurtovic2026minds-5c9} respectively) showing neon ratios above 0.5. For [Ne II]/[Ar II], which we detect for both HK Tau A and B, we recover 3.3 and 2.7 respectively - placing both of them above the 2.5 threshold, which hints at hard X-rays - consistent with the presence of a sufficiently dense inner wind to absorb them.

\subsection{A wind originating from HK Tau B}\label{subsec:discussion_wind}

The 'X' shape of the extended \ch{H2} emission centred in HK Tau B, recovered also in the derived temperature and density structure, is consistent with tracing a wind with a wide opening angle (\ang{77} $\pm$ \ang{13} on the north-west surface and \ang{69} $\pm$ \ang{12} in the south-east one) launching from the upper disc layer. This interpretation is further supported by the difference in the morphology of the \ch{H2} compared to the keplerian \ch{CO}, significantly less extended (see Figure \ref{fig:mom0_threechannels}).

\begin{figure}[htbp]
    \centering
    \includegraphics[width=0.95\linewidth]{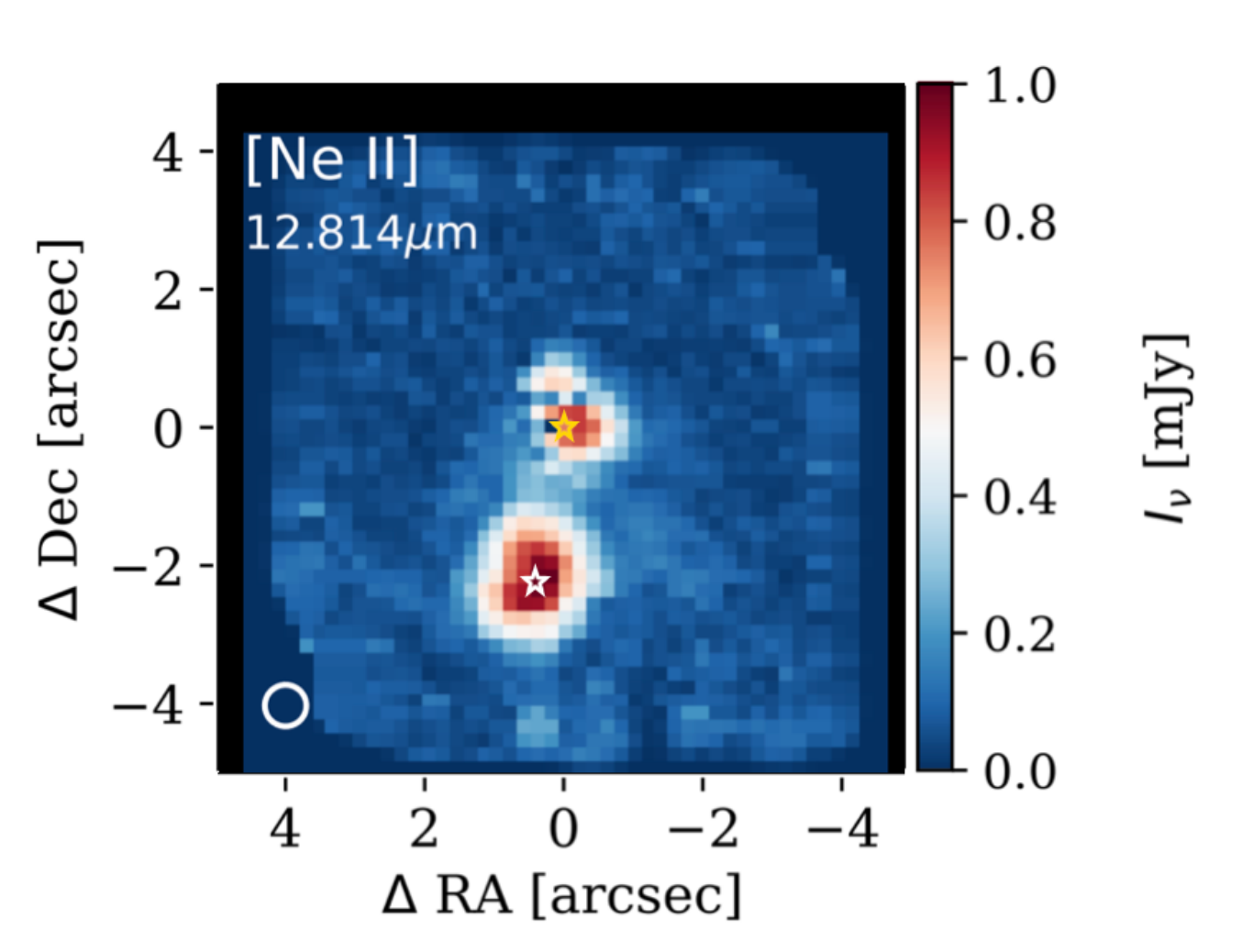}
    \caption{Moment 8 map of the [Ne II] emission around HK Tau A (yellow star) and B (white star).}
    \label{fig:NeII_mom8}
\end{figure}

To investigate the disc- or wind-origin, we inspect the moment 8 map of [Ne II] (Figure \ref{fig:NeII_mom8}). In the surroundings of HK Tau B, [Ne II] has a hourglass shape, perpendicular to the orientation of the continuum; furthermore, it is significantly less extended than the \ch{H2}. This suggests that we are not seeing scattered emission, but rather that the \ch{H2} is tracing the base of a wind originating from HK Tau B. Furthermore, the broad shape of the [Ne II] emission is more compatible with a wind than a high-velocity, collimated jet; this implies that the caveat on the velocity measurement mentioned in Section \ref{subsec:ionised_atoms} on the possibility of a HVC seen as LVC because of inclination effects does not apply. The average \ch{H2} temperature and column density are also consistent with a wind origin. In this interpretation, the large opening angles recovered would suggest that the \ch{H2} is efficiently destroyed in the wind, and therefore only seen near its base. Another argument in favour of \ch{H2} being entrained in a wind is that its temperature is larger than the escape temperature $T_{\rm{esc}}$ \citep{Owen2012}; for example, at the location of the red square in Figure \ref{fig:analysis_T_N_maps},  $T_{\rm{esc}} \simeq 100$ K while the \ch{H2} temperature is a factor 8 larger.

Traditionally, \ch{H2} emission has been preferentially interpreted as a signpost for magnetohydrodynamic disc winds. However, \cite{nakatani2026photoevaporation-d76} have shown that the observed \ch{H2} extended emission can be broadly reproduced by photoevaporation, both in its X-shaped morphology and in the line fluxes. Their radiation hydrodynamics simulations, while not intended to fit any particular source, recovered opening angles and excitation temperatures remarkably similar to the observations of Tau 042021 \citep{arulanantham2024jwst-ecf} and SY Cha \citep{schwarz2025erratum-f59}, suggesting that photoevaporation is indeed a viable explanation. While the opening angles recovered from the theoretical models of \cite{nakatani2026photoevaporation-d76} span between 37 and \ang{50}, significantly smaller than the $\sim$ \ang{70}-\ang{77} we find in HK Tau B, we note that the authors have not conducted a full parameter space exploration and larger opening angles are not ruled out a priori. Furthermore, an increased FUV luminosity is expected to broaden the \ch{H2} dissociation front, therefore widening the opening angle; this is consistent with the the north-western side of the disc having a larger opening angle, as it is illuminated by HK Tau A, as opposed to the south-eastern side. While magnetohydrodynamic winds remain a valid explanation for the observed \ch{H2} emission, we emphasise that a photoevaporative origin cannot be excluded.

\subsection{Lack of PAH emission}\label{subsec:discussion_PAHs}

The lack of PAH emission in HK Tau B contrasts with the recently growing evidence of regular detection of PAH features in edge-on protoplanetary discs. Indeed, while face-on sources show limited to no evidence of PAHs (with a detection rate of $\sim 8 \%$ in discs around T Tauri stars, \citealt{geers2006c2d-96d}), highly inclined systems observed with MIRI-MRS display strong PAH features, especially at 8 and 11\micron; it is the case of Tau 042021 (\ang{88;}, \citealt{arulanantham2024jwst-ecf}), HH 48 NE (\ang{82.3;}, \citealt{sturm2024jwstmiri-3fe}), and T Cha (\ang{73;}, \citealt{bajaj2024jwst-6d4, arun2025when-023}).

PAHs emit in the infrared when exposed to UV radiation, which can explain the lack of features in discs around brown dwarfs \citep{perotti2026minds-913}, very low-mass stars \citep{arabhavi2025minds-bd4}, and low mass T Tauri stars as opposed to intermediate mass T Tauris (PAHs detected in 40 $\%$ of the observed discs, \citealt{valegrd2021what-825}) and Ae/Be Herbig stars (PAHs detected in $70 \%$ of the targets, \citealt{acke2010spitzers-1de}). The lower stellar emission at UV wavelengths combined with depletion of gas-phase PAHs by several orders of magnitude is commonly indicated to be the reason for the small detection rates in discs around low mass stars. On the other hand, why disc inclination would impact the presence of PAH emission is debated. A possible explanation could be localised emission, although the model of \cite{sturm2024jwstmiri-3fe} predicts that the PAHs in HH 48 NE would be observable even in a face-on configuration. Another option is the disc geometry favouring the exposure of PAHs to UV radiation (like the presence of a cavity in the case of HH 48 NE, \citealt{sturm2023-modelingices}), as well as a high UV radiation coming from the companion under certain orientations, or a higher accretion rate. All of these possibilities require further exploration with source-specific modelling. In this context, HK Tau B stands out as the only T Tauri, edge-on source observed at MIRI-MRS wavelengths that does not show signs of PAH emission. We will expand the statistics in a forthcoming paper, which we refer to for a deeper analysis on the presence of PAH features in edge-on T Tauri stars.

\section{Conclusions}\label{sec:conclusions}

In this paper we have presented JWST/MIRI-MRS observations of the binary system HK Tau, which is composed of a low-inclination primary and close-to-edge-on secondary. We have analysed the mid infrared spectrum of both sources and discussed their molecular and atomic emission features, as well as ice absorption bands visible in the secondary. Our main results are the following:

\begin{enumerate}

    \item While HK Tau A shows a line- (especially \ch{CO2}-) rich spectrum, HK Tau B is stunningly line poor - except for atomic ions and \ch{H2} emission. This is likely a consequence of the geometrical configuration, as we have shown that sources with inclinations > \ang{75} are expected to lose molecular emission signatures, while maintaining ionised atomic lines that originate in the upper layers of the disc;

    \item The high inclination of HK Tau B allows detection of ice absorption bands: we confidently detect \ch{H2O} at 6 and 13.6\micron, \ch{CO2} at 15\micron, and \ch{NH4+} at 6.85\micron. We tentatively detect \ch{CH4} at 7.6\micron. 
    
    \item Under the assumption of coeval sources and similar disc evolution, the complementary inclination of the two sources allows to simultaneously probe the solid- and gas-phase content of the system. The MIR spectra suggest a similar chemical composition with significant water and \ch{CO2};

    \item The \ch{H2} emission is extended in both sources. The 'X'-shaped emission centred in B shows thermochemical conditions compatible with a wind (photoevaporative or MHD) origin; the asymmetric, wide opening angle (\ang{77} and \ang{69} for the upper and lower surfaces respectively) suggests that the outflowing \ch{H2} is efficiently destroyed in the wind, and therefore only seen near the wind base;

    \item We detect [Ne II] and [Ar II] in both sources, as well as [Ne III] in HK Tau B. The hourglass shape of the [Ne II] emission further points in the direction of a wind irradiating from HK Tau B. The line ratios place HK Tau A and B in the X-ray dominated irradiation regime;

    \item Unlike all other T Tauri, edge-on protoplanetary discs observed with MIRI-MRS so far, HK Tau B does not show any sign of PAH emission. Larger samples of edge-on discs are needed to put this result in context.

\end{enumerate}

The peculiar configuration of HK Tau offer the unprecedented opportunity to have a simultaneous view on two coeval, almost identical mass pre-main sequence objects with complementary inclinations. The high angular resolution of JWST/MIRI-MRS has allowed to separate the contributions of the two sources, shedding light on molecular diversity, atomic emission lines, and wind tracers. With this work, we make a step forward in the JWST/MIRI characterisation of both multiple systems and edge-on discs, and highlight the need for larger statistical samples to interpret our results in a population framework.

\vspace{0.3cm}
%-----------------------------------------------------------------

\tiny{\noindent\textit{Acknowledgements.} We thank an anonymous referee for constructive input that has helped us improve the quality and clarity of the manuscript. AS thanks Jennifer Bergner, Gabriele Cugno, Jenny Frediani, Ryohei Nakatani, and Tushar Suhasaria for interesting discussions.}

\tiny{This work is based on observations made with the NASA/ESA/CSA James Webb Space Telescope. The data were obtained from the Mikulski Archive for Space Telescopes at the Space Telescope Science Institute, which is operated by the Association of Universities for Research in Astronomy, Inc., under NASA contract NAS 5-03127 for JWST. These observations are associated with program \#1282. The following National and International Funding Agencies funded and supported the MIRI development: NASA; ESA; Belgian Science Policy Office (BELSPO); Centre Nationale d’Etudes Spatiales (CNES); Danish National Space Centre; Deutsches Zentrum fur Luft- und Raumfahrt (DLR); Enterprise Ireland; Ministerio De Econom\'ia y Competividad; Netherlands Research School for Astronomy (NOVA); Netherlands Organisation for Scientific Research (NWO); Science and Technology Facilities Council; Swiss Space Office; Swedish National Space Agency; and UK Space Agency.}

\tiny{This paper makes use of the following ALMA data: ADS/JAO.ALMA$\#$2016.1.00460.S, ADS/JAO.ALMA$\#$2018.1.00771.S. ALMA is a partnership of ESO (representing its member states), NSF (USA) and NINS (Japan), together with NRC (Canada), NSTC and ASIAA (Taiwan), and KASI (Republic of Korea), in cooperation with the Republic of Chile. The Joint ALMA Observatory is operated by ESO, AUI/NRAO and NAOJ. The National Radio Astronomy Observatory is a facility of the National Science Foundation operated under cooperative agreement by Associated Universities, Inc.}

\tiny{G.P. gratefully acknowledges support from the Carlsberg Foundation, grant CF23-0481 and from the Max Planck Society. T.H. acknowledges support from the European Research Council under the Horizon 2020 Framework Program via the ERC Advanced Grant Origins 83 24 28. E.v.D. acknowledges support from the ERC grant 101019751 MOLDISK and the Danish National Research Foundation through the Center of Excellence ``InterCat'' (DNRF150). A.D.S., M.T., and M.V. acknowledge support from the ERC grant 101019751 MOLDISK. I.K., A.M.A., and E.v.D. acknowledge support from grant TOP-1 614.001.751 from the Dutch Research Council (NWO). A.C.G. acknowledges support from PRIN-MUR 2022 20228JPA3A “The path to star and planet formation in the JWST era (PATH)” funded by NextGeneration EU and by INAF-GoG 2022 “NIR-dark Accretion Outbursts in Massive Young stellar objects (NAOMY)” and Large Gran INAF-2024 “Spectral Key fea-tures of Young stellar objects: Wind-Accretion LinKs Explored in the infraRed (SKYWALKER)”. V.C. acknowledges funding from the Belgian F.R.S.-FNRS. T.K. acknowledges support from STFC Grant ST/Y002415/1. L.M.S. has received funding from the European Research Council (ERC) under the European Union’s Horizon 2020 research and innovation programme (PROTOPLANETS, grant agreement No. 101002188). B.T. is a Laureate of the Paris Region fellowship program, which is supported by the Ile-de-France Region and has received funding under the Horizon 2020 innovation framework program and Marie Sklodowska-Curie grant agreement No. 945298.}

\bibliographystyle{aa}
\bibliography{bibliography}

\begin{appendix}
\normalsize

\section{Comparison with Spitzer spectrum}\label{appendix:comparison_Spitzer}

\begin{figure}[htbp]
    \centering
    \includegraphics[width=\linewidth]{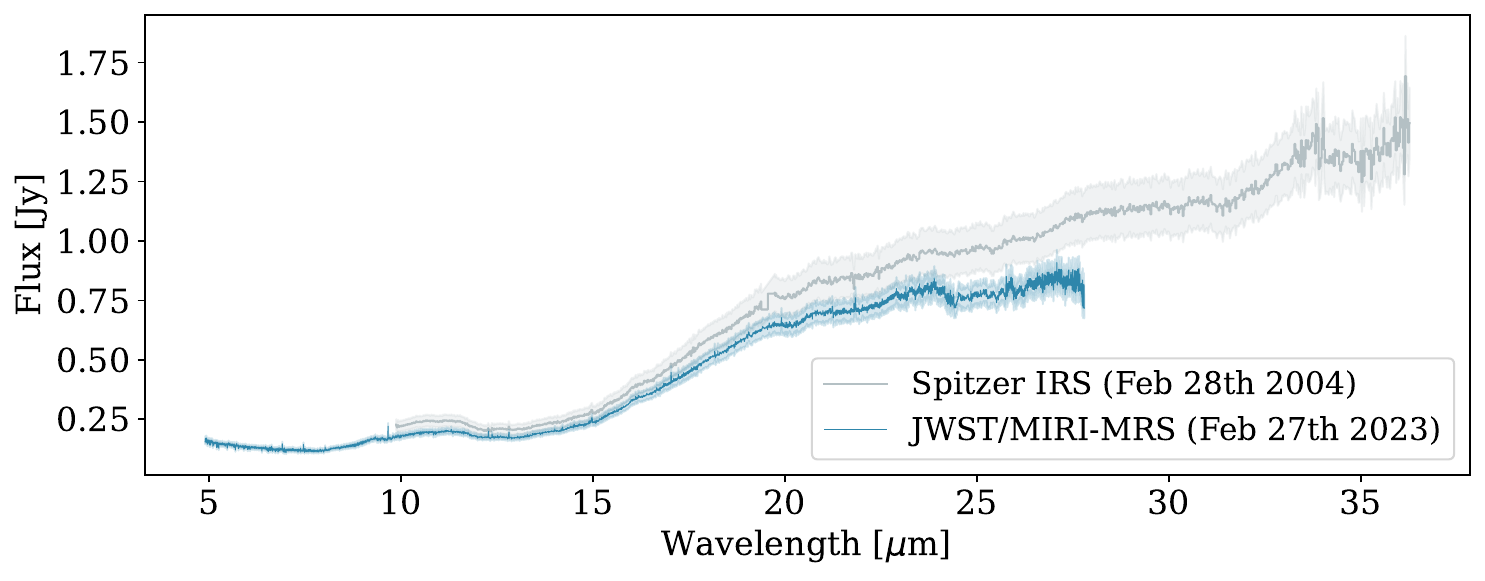}
    \caption{Comparison of the total spectrum of the HK Tau system obtained with \textit{Spitzer} IRS (grey) and JWST/MIRI-MRS (blue), taken 19 years apart. The combined spectrum is dominated by the primary component. The shaded regions represent the reported spectrophotometric accuracy for both instruments.} 
    \label{fig:Spitzer_comparison}
\end{figure}

The HK Tau system has been observed with the \textit{Spitzer} InfraRed Spectrograph (IRS), 19 years before the MINDS observations. Figure \ref{fig:Spitzer_comparison} shows the JWST/MIRI-MRS spectrum compared with the archival IRS spectrum\footnote{Reduced data credits: Klaus Pontoppidan (available for download at \url{https://www.stsci.edu/~pontoppi/\#data}).}. While we are able to disentangle the two components of the binary system in MIRI-MRS, this is not the case for IRS, hence we compare the fluxes of the combination of HK Tau A and HK Tau B. We note that the flux of HK Tau A is approximately two orders of magnitude larger than that of B, therefore the spectra in Figure \ref{fig:Spitzer_comparison} are dominated by the primary component.

The \textit{Spitzer} flux is systematically higher than that obtained with JWST, which could in principle suggest variability. Mid-infrared variability in Class II discs, observed when comparing IRS and MIRI-MRS spectra, usually shows one of two behaviours - a consistently stronger, or weaker, flux across the whole wavelength range (attributed to a change in the incident flux from the central object) and a ‘seesaw’ pattern, with the emission varying inversely at wavelengths shorter and longer than a pivotal point (see, e.g., \citealt{espaillat2011spitzer-8df, perotti2026minds-913}). As we do not have access to the IRS flux at wavelengths shorter than 10\micron, we cannot rule out any of the two behaviours. Furthermore, once we account for the expected calibration errors - based on the spectrophotometric accuracy of both instruments: between 2 and 10$\%$ for IRS \citep{furlan2006survey-4d1, watson2009crystalline-1e6} and $5.6 \pm 0.7 \%$ for MIRI-MRS \citep{argyriou2023jwst-f58} -, the fluxes overlap for wavelengths shorter than $\sim$ 22\micron; at longer wavelengths instead, the discrepancy is larger - although still moderate. A detailed radiative transfer modelling, and potentially more observed epochs, would be needed to fully appreciate the cause of this discrepancy.

\newpage

\section{Results of the slab model fit}\label{appendix:slab_results}

\begin{table}[h]
    \centering

    \caption{Parameters of the best fit slab model for the continuum-subtracted spectrum of HK Tau A.}
    
    \begin{tabular}{c|c|c|c}

    \hline

    \rule{0pt}{2.5ex} Molecule & $T$ [K] & $N$ [cm$^{-2}$] & $R_{slab}$ [au] \\[1ex]

    \hline

    \rule{0pt}{2.5ex} \ch{H2O} & 725 & $2.15 \times 10^{18}$ & 0.13 \\
    \ch{CO2} & 350 & $4.64 \times 10^{17}$ & 0.01 \\
    \ch{HCN} & 175 & > $4.64 \times 10^{14}$ & < 1 \\
    \end{tabular}
    \tablefoot{The fit was performed in the 13.5-16.2\micron\ region (Figure \ref{fig:radexpy_A}) . We do not report the fit results for \ch{OH} as it is very likely not in LTE in T Tauri discs, and therefore the retrieved parameters (especially the temperature) are not representative of the physical conditions of the gas (see, e.g., \citealt{tabone2021oh-443, tabone2024oh-6af}).}
    \label{tab:appendix_fit}
\end{table}

\section{Rotational diagram constants and moment 0 maps}\label{appendix:RD_and_maps}

\begin{table}[htbp]

    \centering

    \caption{\ch{H2} rotational transition properties.}\label{tab:H2_transitions}
    
    \begin{tabular}{c c c c c}
    
    \hline
    \hline

    \rule{0pt}{2.5ex} Transition & $\lambda$ ($\mu$m) & $E_u$ (K) & $g_u$ & $A_u$ (s$^{-1}$) \\[1ex]

    \hline

    \rule{0pt}{2.5ex}S(1) & 17.0348 & 1015.1 & 21 & $4.761 \times 10^{-10}$ \\
    S(2) & 12.2786 & 1681.6 & 9 & $2.775 \times 10^{-9}$ \\
    S(3) & 9.66491 & 2503.7 & 33 & $9.836 \times 10^{-9}$ \\
    S(4) & 8.02504 & 3474.5 & 13 & $2.643 \times 10^{-8}$ \\
    S(5) & 6.90951 & 4586.1 & 45 & $5.879 \times 10^{-8}$ \\
    S(6) & 6.10856 & 5829.8 & 17 & $1.142 \times 10^{-7}$ \\
    S(7) & 5.51118 & 7196.7 & 57 & $2.001 \times 10^{-7}$ \\
    S(8) & 5.05306 & 8677.1 & 21 & $3.236 \times 10^{-7}$ \\

    \end{tabular}

    \hspace{1.5cm}

    \tablefoot{$\lambda$: the transition wavelength; $E_u$: energy of the upper level measured from the ground $\nu = 0, J = 0$ level; $g_u$: statistical weight of the upper level; $A_{\mathrm{u}}$: electric quadrupole emission probability. $E_u$, $g_u$, $A_u$ from \cite{roueff2019full-9c7}.}

    \label{tab:analysis_rotvi_constants}

\end{table}

\vspace{1ex}

\section{Complete moment 0 \ch{H2} maps}\label{appendix_fullmom0}

\begin{figure}[htbp]
    \centering
    \includegraphics[width=0.96\linewidth]{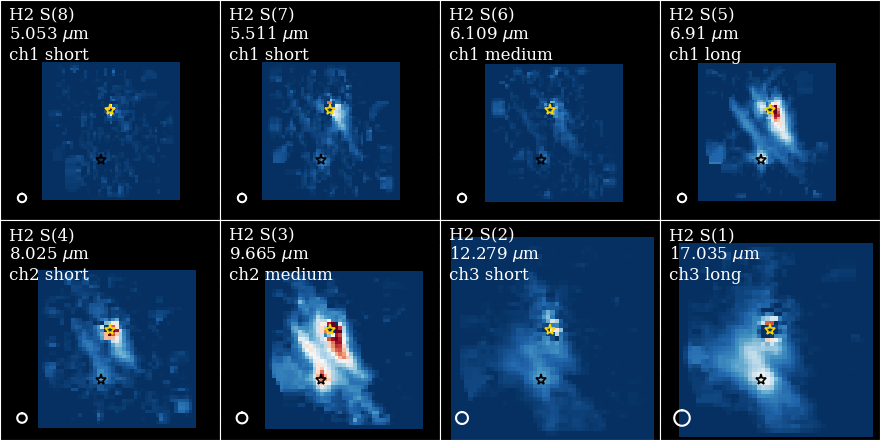}
    \caption{Same as Figure \ref{fig:mom0_threechannels} including all \ch{H2} extended lines within the MIRI-MRS wavelength range.}
    \label{fig:mom0_complete}
\end{figure}

\end{appendix}

\end{document}